\documentclass[aps,prd,reprint,groupedaddress]{revtex4-1}

\usepackage{graphicx}
\usepackage{amsmath}
\usepackage{hyperref}

\begin{document}


\title{Exploring nucleon spin structure through neutrino neutral-current interactions in MicroBooNE}


\author{K. Woodruff,\; for the MicroBooNE collaboration}
\affiliation{New Mexico State University}


\date{\today}

\begin{abstract}
The net contribution of the strange quark spins to the proton spin, $\Delta s$,
can be determined from neutral current elastic neutrino-proton interactions at
low momentum transfer combined with data from electron-proton scattering. The
probability of neutrino-proton interactions depends in part on the axial form factor,
which represents the spin structure of the proton and can be separated into its
quark flavor contributions.  Low momentum transfer neutrino neutral current
interactions can be measured in MicroBooNE, a high-resolution liquid argon time
projection chamber (LArTPC) in its first year of running in the Booster
Neutrino Beamline at Fermilab. The signal for these interactions in MicroBooNE
is a single short proton track. We present our work on the automated
reconstruction and classification of proton tracks in LArTPCs, an important
step in the determination of neutrino-nucleon cross sections and the
measurement of $\Delta s$.
\end{abstract}

\pacs{}

\maketitle

\section{Introduction \label{intro}}
The structure of a nucleon is more interesting than the three familiar up and
down valence quarks. These three quarks only account for a small percent of the
nucleon mass. The gluons that bind the quarks split into quark-antiquark pairs
of up, down, and strange flavor. The remainder of the nucleon mass is carried
by this quark-gluon sea. The structure of the sea and how its elements combine
with the valence quarks to give the nucleon its measured structure is not
precisely known.

The net spin of the proton comes from a combination of the spin and orbital
momentum of the quarks and gluons. The net contribution from the spin of
strange quarks and antiquarks, $\Delta s$, is defined as
\begin{equation*}
  \Delta s = \int_0^1 \Delta s(x) \, dx
\end{equation*}
\begin{equation*}
  \Delta s(x) = \sum_{r=\pm 1} r[s^{(r)}(x) + \bar{s}^{(r)}(x)] \,,
\end{equation*}
where $s(\bar{s})$ is the spin-dependent parton distribution function of the
strange (anti)quark, $r$ is the helicity of the quark relative to the proton
helicity and $x$ is the Bjorken scaling variable~\cite{Alberico01}.  In the
static quark model this value is zero.

In the 1980s the European Muon Collaboration~\cite{Ashman89} and several
subsequent experiments found that the Ellis-Jaffe Sum Rule was violated in
polarized, charged-lepton, inclusive, deep inelastic scattering (DIS). The
Ellis-Jaffe sum rule~\cite{Ellis74} assumes that SU(3) flavor symmetry is valid and that
$\Delta s = 0$. For the results to be consistent with exact SU(3) flavor
symmetry, $\Delta s$ must be \textit{negative}. Follow-up measurements using
semi-inclusive deep inelastic scattering have been consistent with $\Delta s =
0$, but these determinations of $\Delta s$ are highly dependent on the
fragmentation functions used~\cite{Aidala12}.

An independent determination of $\Delta s$ can be made using neutral-current
(NC) elastic neutrino-proton scattering. The NC elastic cross section depends
directly on $\Delta s$ and no assumptions about SU(3) flavor symmetry or
fragmentation functions are needed.

\section{Elastic neutrino-proton scattering \label{elnup}}
The elastic lepton-nucleon scattering cross section depends on the axial,
electric, and magnetic form factors which represent the finite structure of the
nucleon. The axial form factor, $G_A$, represents the spin structure, and the
electric and magnetic form factors, $G_E$ and $G_M$, represent the electric and
magnetic structure, respectively.

\subsection{Neutral-current elastic scattering \label{ncel}}
The NC elastic neutrino-proton cross section~\cite{Alberico01} can be written as
\begin{equation*}
  \begin{split}
    \left(\frac{d\sigma}{dQ^2}\right)_{\nu}^{NC} &=\frac{G_F^2}{2\pi} \left[
    \frac{1}{2}y^2(G_M^{NC})^2 \right. \\
    &\left. +\left(1-y-\frac{M}{2E}y \right)
    \frac{(G_E^{NC})^2+\frac{E}{2M}y(G_M^{NC})^2}{1+\frac{E}{2M}y}
    \right.\\
    &+ \left. \left(\frac{1}{2}y^2 + 1 - y + \frac{M}{2E}y
    \right)(G_A^{NC})^2 \right. \\
    &+ \left. 2y \left(1-\frac{1}{2}y \right)
    G_M^{NC}G_A^{NC} \right] \,,
  \end{split}
\end{equation*}
where $G_F$ is the Fermi constant, $M$ is the mass of the nucleon, $E$ is the
neutrino energy, and $Q^2$ is the momentum transfer.

The neutral-current form factors, $G_A^{NC}$, $G_E^{NC}$, and $G_M^{NC}$, are
functions of $Q^2$ and can all be written as a linear combination of the
individual quark contributions
\begin{equation*}
  \begin{split}
    G_{E,M}^{NC,p}(Q^2) &= \left(1-\frac{8}{3}\textrm{sin}^2\theta_W\right)G_{E,M}^u(Q^2) \\
    &+ \left(-1+\frac{4}{3}\textrm{sin}^2\theta_W\right)G_{E,M}^d(Q^2) \\
    &+ \left(-1 + \frac{4}{3}\textrm{sin}^2\theta_W\right)G_{E,M}^s(Q^2) \\
    G_A^{NC,p}(Q^2) &= \frac{1}{2}\left[-G_A^u(Q^2) + G_A^d(Q^2) + G_A^s(Q^2) \right] \,.
  \end{split}
\end{equation*}
The up, down, and strange quark contributions to the electric and magnetic form
factors of the proton have been determined in a world-wide measurement program
of elastic electron-proton scattering using hydrogen targets and quasi-elastic
electron-nucleon scattering using light nuclear targets (specifically deuterium
and helium)~\cite{Armstrong12,Cates11}.

We plan to measure the ratio of the neutral-current elastic cross section to
the charged-current elastic cross section. The charged-current (CC) elastic
cross section does not depend on $\Delta s$, but it is better known than the NC
elastic cross section. Taking the ratio of the two cross sections reduces
systematic uncertainty on our measurement due to the beam flux, detector
efficiency, and nuclear effects and final state interactions in argon nuclei.

\subsection{Axial form factor \label{axial}}
At the limit when the momentum transfer ($Q^2$) goes to zero, the quark
contributions to the axial form factor become the net contribution of
individual quark spin to the proton spin,
\begin{equation*}
  G_A^q(Q^2 = 0) = \Delta q \qquad (q=u,d,s) \,,
\end{equation*}
so that
\begin{equation*}
  G_A^{NC}(Q^2 = 0) = \frac{1}{2}(-\Delta u + \Delta d + \Delta s) \,.
\end{equation*}
The difference of the up and down spin contributions, $\Delta u - \Delta d$, is
proportional to the axial vector coupling constant $g_A$ measured in hyperon
$\beta$ decay~\cite{Olive16}, therefore a measurement of $G_A^{NC}$ can
determine $\Delta s$.

\subsection{Experimental measurement}

The final state of an NC elastic neutrino-proton interaction consists of a
neutrino and a proton. Since it isn't possible to detect the outgoing neutrino,
the signal is a single proton track. In order to extrapolate the axial form
factor to zero, we need to detect very low energy protons. The kinematics of
the interaction are determined entirely by the proton kinetic energy, $T_P$,
\begin{equation*}
  Q^2 = 2T_P M \,.
\end{equation*}

Previous measurements~\cite{Ahrens86,Garvey93,Aguilar10} have been able to
resolve final state protons down to a kinetic energy of $T\sim~240$~MeV which
corresponds to a momentum transfer of $Q^2=0.45$~GeV$^2$. These measurements
also found $\Delta s < 0$, but the results are highly dependent on the choice
of the axial form factor $Q^2$ dependence. To extract $\Delta s$, $G_A^s$ must
be extrapolated to $Q^2 = 0$. Detecting events with lower momentum transfer
would lessen the dependence on the choice of the model.

We estimate that MicroBooNE can detect NC elastic events down to a minimum of
$Q^2 \sim 0.08$ GeV$^2$.  The momentum transfer is determined by the kinetic
energy of the proton in NC elastic interactions. MicroBooNE can detect protons
with a track length of at least 1.5 cm which corresponds to a kinetic energy of
$\sim$40 MeV in liquid argon giving $Q^2 \sim 0.08$~GeV$^2$.

\section{MicroBooNE \label{uboone}}
\begin{figure}
  \includegraphics[scale=0.15]{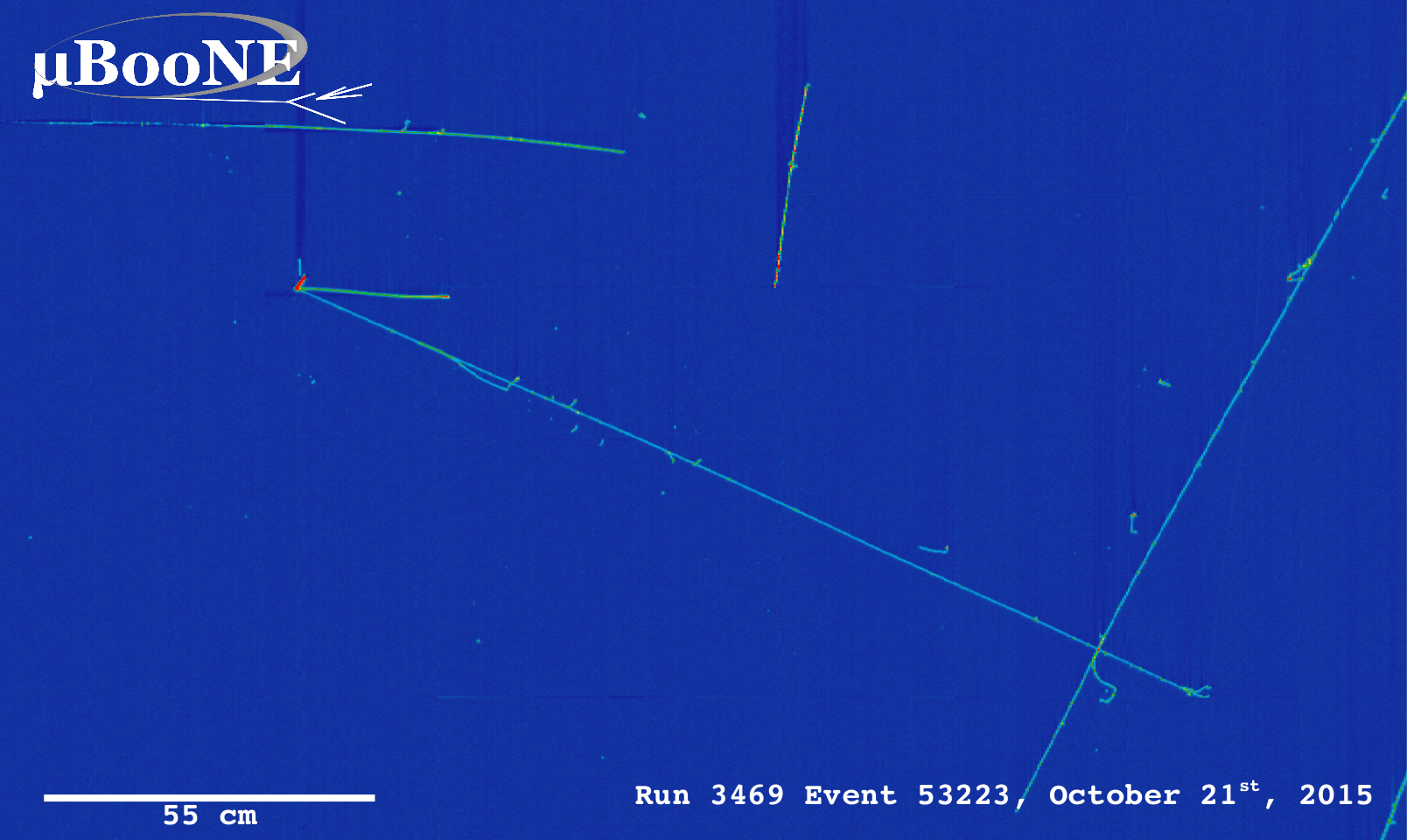}
  \caption{A neutrino interaction in the MicroBooNE detector. This is a
  charged-current, muon-neutrino event with a long muon track, a charged pion
  track, and a short proton track coming from the interaction vertex.
  \label{fig:event}}
\end{figure}
\begin{figure}
  \includegraphics[scale=0.43]{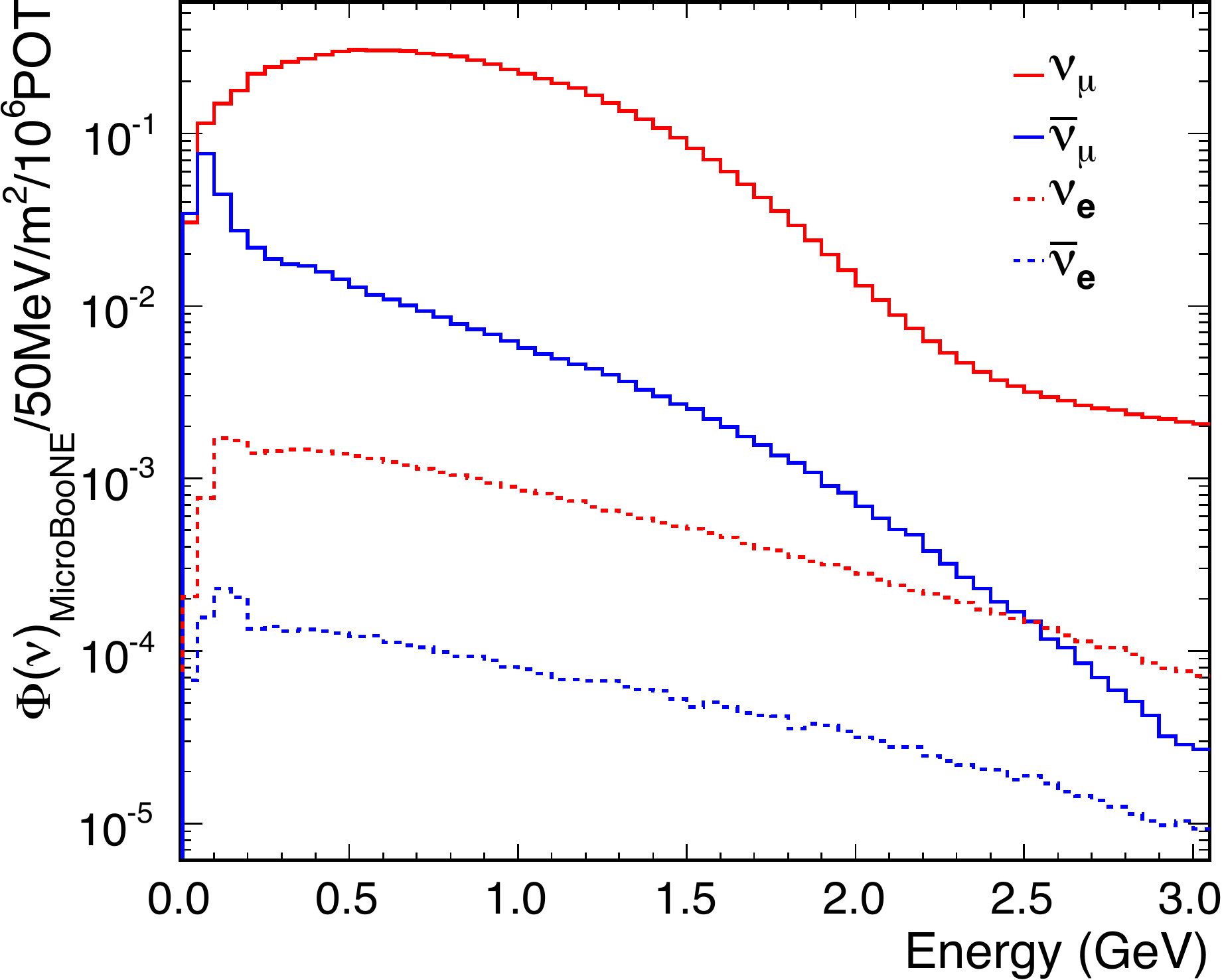}
  \caption{The estimated Booster Neutrino Beam flux at MicroBooNE. \label{fig:bnbflux}}
\end{figure}

The MicroBooNE detector~\cite{UBdetector} is a liquid-argon time projection
chamber (TPC) located in the Booster Neutrino Beam at Fermilab. MicroBooNE is a
high-resolution detector designed to be able to accurately identify low-energy
neutrino interactions. It began taking data in October of 2015. Figure~\ref{fig:event}
shows an example neutrino interaction in MicroBooNE.

\subsection{The beam}
The Booster Neutrino Beam (BNB)~\cite{Aguilar09} is produced by protons from
the Booster synchrotron incident on a beryllium target. The proton beam has a
kinetic energy of 8 GeV, a repetition rate of 5 Hz, and an intensity of
$5\times10^{12}$ protons-per-spill. Secondary pions and kaons decay producing
neutrinos with an average energy of $\sim$800 MeV. The estimated BNB flux is
shown in Fig.~\ref{fig:bnbflux}. MicroBooNE has received $3.6\times 10^{20}$
protons-on-target in its first year of running.

\subsection{The detector}
The MicroBooNE TPC~\cite{UBdetector} has an active mass of 89 tons of liquid
argon. It is 10 meters long in the beam direction, 2.3 meters tall, and 2.5
meters in the electron drift direction. It takes 2.3 ms for electrons to drift
across the full width of the TPC at the operating electric field of 273 V/cm.
Events are read out on three anode wire planes with 3 mm spacing. In addition
to the TPC, there is a light collection system which consists of 32 8-inch PMTs
with nanosecond timing resolution.  The PMTs determine the initial time of the
interaction to help with cosmic rejection. In order for an event to be read
out, there must be an optical signal within a 23 $\mu$s window around the BNB
spill.

\subsection{The effect on $\Delta s$ uncertainty}
Global fits to electron-proton and neutrino-proton elastic scattering data have
found $\Delta s = -0.30 \pm 0.42$~\cite{Pate13}. Based on data from a simulation
of the MicroBooNE detector and the BNB beam, the uncertainty on the global fit
to $\Delta s$ is estimated to decrease by a factor of ten when including
MicroBooNE data.

\section{Automated event selection \label{evtsel}}
\begin{figure}
  \includegraphics[scale=0.22]{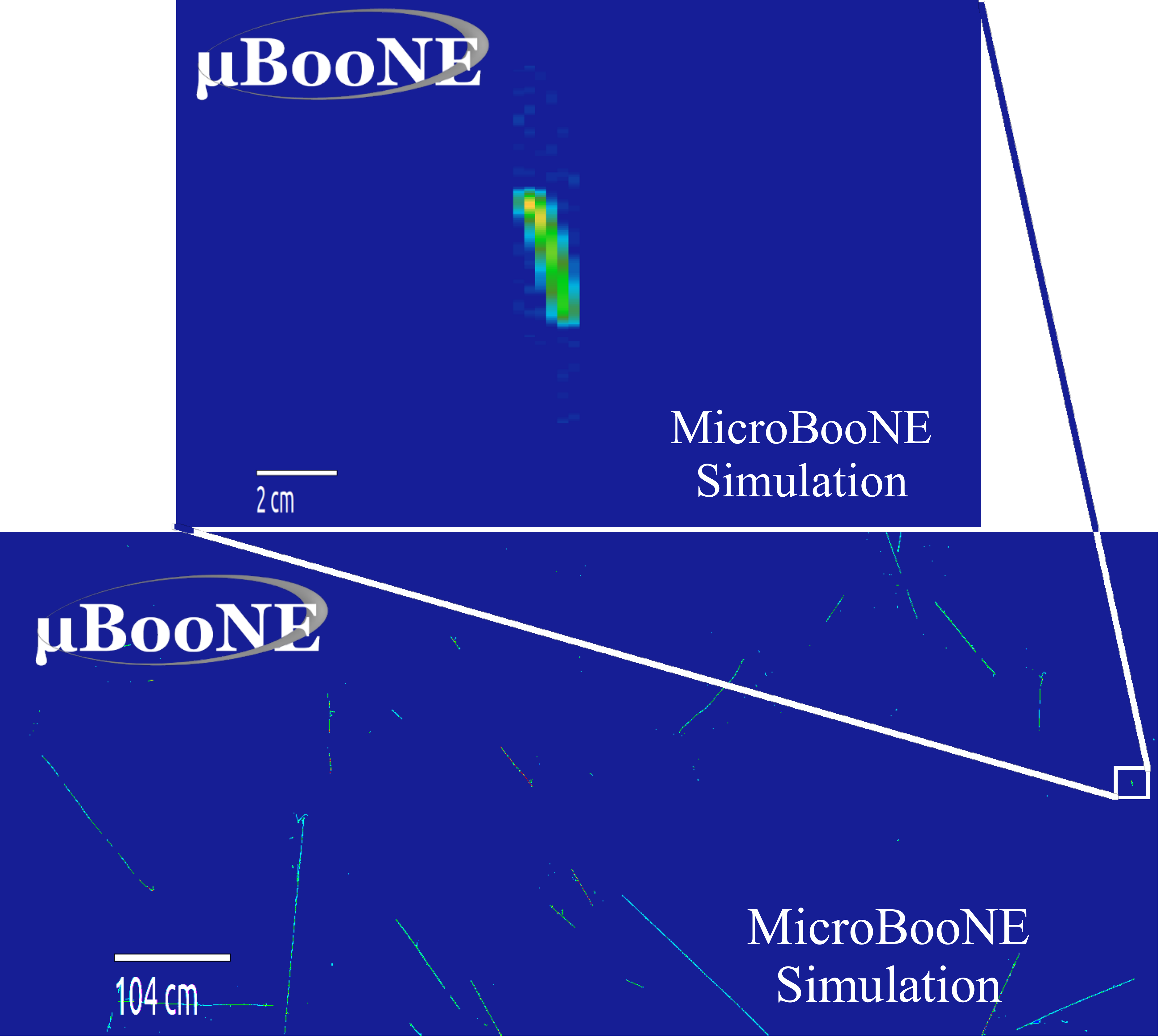}
  \caption{2D event display of a simulated neutral-current elastic event in
  MicroBooNE that was classified as a proton. The top image is a close-up event
  display of the simulated proton track. The bottom image shows the side view
  of the entire MicroBooNE TPC. All of the additional tracks are from cosmic
  rays.  \label{fig:mcevd}}
\end{figure}

MicroBooNE is close to the surface of the Earth, which results in a large
cosmic ray background. Each triggered event is read out for 4.8 ms
(approximately twice the electron drift time), and there are an average of
twelve cosmic muon tracks per readout frame~\cite{UBcosmic}.  This can be seen
in the bottom image in Fig.~\ref{fig:mcevd}. In addition, there are
approximately five times as many event triggers caused by cosmic rays
coincident with the BNB spill than actual neutrino interactions. During
MicroBooNE's three year run, we expect to have $\sim$200,000 neutrino
interactions and $\sim$1,000,000 cosmic interactions.  This means that
automated neutrino event reconstruction and identification algorithms are
required. These algorithms are currently being developed for liquid argon TPCs.

\subsection{Track reconstruction in LArSoft \label{larreco}}
\begin{figure}
  \includegraphics[scale=0.3]{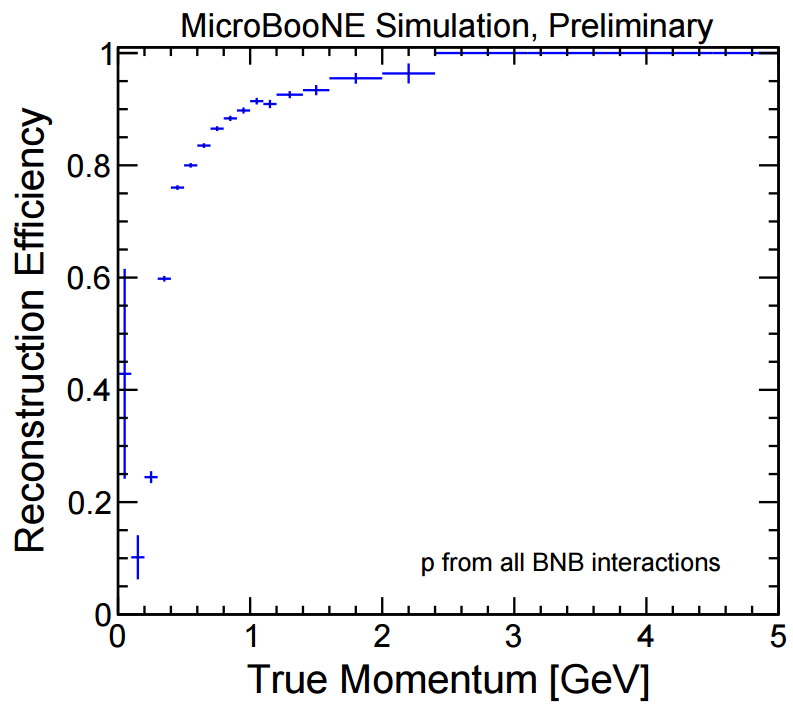}
  \caption{Reconstruction efficiency of the Pandora algorithm for the proton
  with the most hits in all simulated BNB interactions, shown as a function of
  their true momentum~\cite{UBpandora}. \label{fig:paneff}}
\end{figure}

Track reconstruction is handled in the Liquid Argon Software framework
(LArSoft)~\cite{Church13}. The three main stages of reconstruction in LArSoft are hit finding,
track finding, and event identification.

One-dimensional hits are found by fitting Gaussian functions to noise-filtered~\cite{UBnoise}
waveforms that are read out from the anode wires in the TPC. This is done for
all of the wires on all three of the planes. The result is a two-dimensional
image for each of the three wire planes, where the two dimensions are wire
number and time. These 2D hits are used as inputs to the Pandora Software
Development Kit~\cite{UBpandora}.  Pandora contains pattern recognition algorithms that have
been optimized to reconstruct tracks from neutrino interactions in liquid argon
TPCs at the BNB energy range.  The Pandora algorithms take a set of hits and
reconstruct neutrino interaction vertices.

Neutral-current elastic interactions are the most difficult to detect
automatically because there is only one visible particle coming from the
interaction vertex. There is no unique topology separating these events from
the cosmic background. The current reconstruction efficiency for tracks from NC
elastic proton events in simulation is approximately 0.5, and this number is
rapidly improving. The reconstruction efficiency for the proton with the most
hits in all BNB interactions is shown as a function of simulated proton
momentum in Fig.~\ref{fig:paneff}. Once tracks are reconstructed we attempt to
identify the type of particle and interaction that produced them. In the NC
elastic case, we want to specifically select proton tracks.

\subsection{Proton track identification \label{protonid}}

\subsubsection{Gradient decision tree boosting}
To identify proton tracks, we use a gradient-boosted decision tree classifier.
We chose to use decision trees because they are easily interpretable and the
inputs can be a mix of numeric and categorical variables. Below is a short
description of gradient tree boosting. A more detailed description can be found
in the documentation for the XGBoost\cite{Chen16} software library that was
used. 

A decision tree can be thought of as a series of if/else statements that
separate a data set into two or more classes. The goal of each cut is to
increase the information gain. For numerical variables any cut value can be
selected by the tree.  At each node of the tree, a split is chosen to maximize
information gain until a set level of separation is reached.  At the terminus
of the series of splits, called a leaf, a class is assigned.

Two weaknesses of decision trees are their tendency to over fit the training
data and the fact that the output is a class label and not a probability.
Gradient-boosting addresses both of these issues by combining many weak
classifiers into a strong one. Each weak classifier is built based on the error
of the previous one. For a given training set, whenever a sample is classified
incorrectly by a tree, that sample is given a higher importance when the next
tree is being created.  Mathematically, each tree is training on the gradient
of the loss function. After all of the trees have been created, each tree is
given a weight based on its ability to classify the training set, and the
output of the gradient-boosted decision tree classifier is the probability that
a sample is in a given class.

\subsubsection{The decision tree model}
We created a multi-class gradient-boosted decision tree classifier, using the
XGBoost software library, to separate five different track types: any proton
track, muons or pions from BNB neutrino interactions, tracks from
electromagnetic showers from BNB interactions, and any non-proton track
produced by a cosmic ray interaction. The classifier takes reconstructed track
features as input and outputs a probability of the track having been produced
by each of the given particle types. The reconstructed features are based on
the track's geometric, calorimetric, and optical properties.

The training data that we use to make the decision trees comes from Monte Carlo
simulation. The BNB interactions are simulated using the GENIE neutrino
generator~\cite{Andreopoulos09}, and cosmic interactions are simulated using
the CORSIKA cosmic ray generator~\cite{Heck98}. The particles generated by
GENIE and CORSIKA are passed to Geant4~\cite{Agostinelli02} where they are
propagated through a simulated MicroBooNE detector. For training and testing of
the trees we only use tracks that were reconstructed in LArSoft.

\begin{figure}
  \includegraphics[scale=1.]{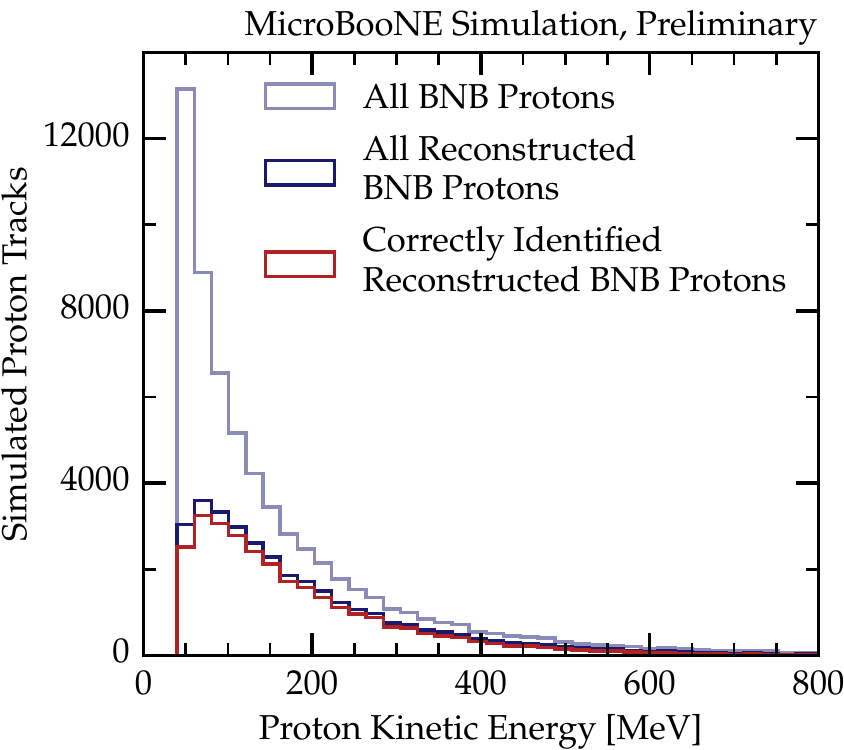}
  \caption{Number of simulated proton tracks as a function of true simulated
  kinetic energy is shown. The light blue line shows the total number of
  protons from simulated BNB neutrino interactions. The dark blue line shows
  the total number of those tracks that were reconstructed with the Pandora
  algorithms. The red line shows the subset of the reconstructed tracks that
  are classified as protons by the boosted decision trees. \label{fig:effvke}}
\end{figure}

\begin{figure}
  \includegraphics[scale=1.]{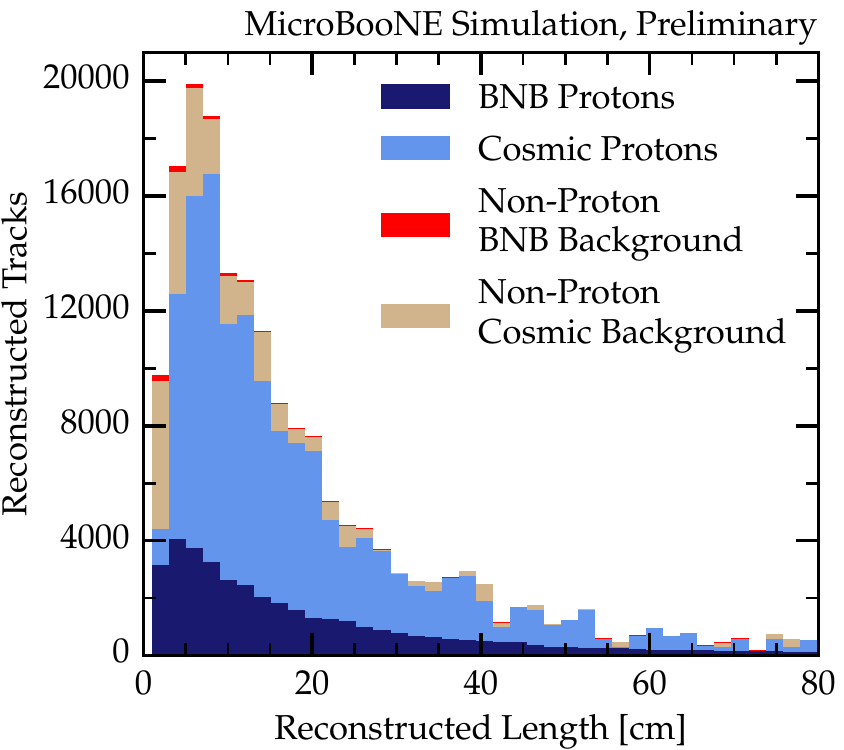}
  \caption{Breakdown of the simulated particle types that are classified as
  protons by the boosted decision trees as a function of reconstructed track
  length. The blue filled area shows all simulated protons, both cosmic and
  neutrino-induced, and the dark blue line shows the protons from simulated BNB
  neutrino interactions. The tan filled area shows all other simulated cosmic
  tracks that are classified as protons, and the red filled area shows all
  other tracks from simulated BNB neutrino interactions that are classified as
  protons. \label{fig:bkgd}}
\end{figure}

\begin{figure}
  \includegraphics[scale=1.]{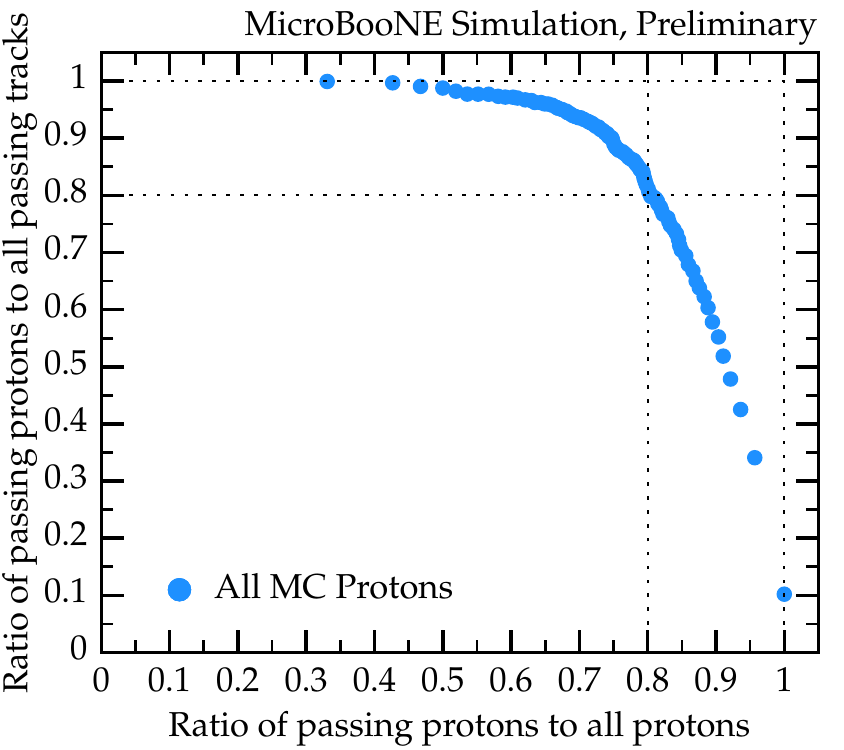}
  \caption{The efficiency versus the purity of simulated protons selected by
    the boosted decision tree classifier for a series of proton probability
      cuts between zero and one.  \label{fig:effvpur}}
\end{figure}

Of the reconstructed test tracks that were input to the classifier, 84\% of the
protons from simulated neutrino interactions, and 63\% of the protons from
simulated cosmic interactions were classified correctly as protons.
Figure~\ref{fig:effvke} shows the protons from simulated neutrino interactions
as a function of proton kinetic energy. Of the reconstructed test tracks that
were classified as protons, 89\% were true simulated protons (22\% neutrino
    induced protons and 67\% cosmic induced protons). Figure~\ref{fig:bkgd}
shows the breakdown of track types that are classified as protons. To maximize
efficiency or purity we can require a lower or higher proton probability from
the classifier. Figure~\ref{fig:effvpur} shows the efficiency versus purity for
different proton probability cuts in the range from zero to one.


\begin{figure}
  \includegraphics[scale=0.225]{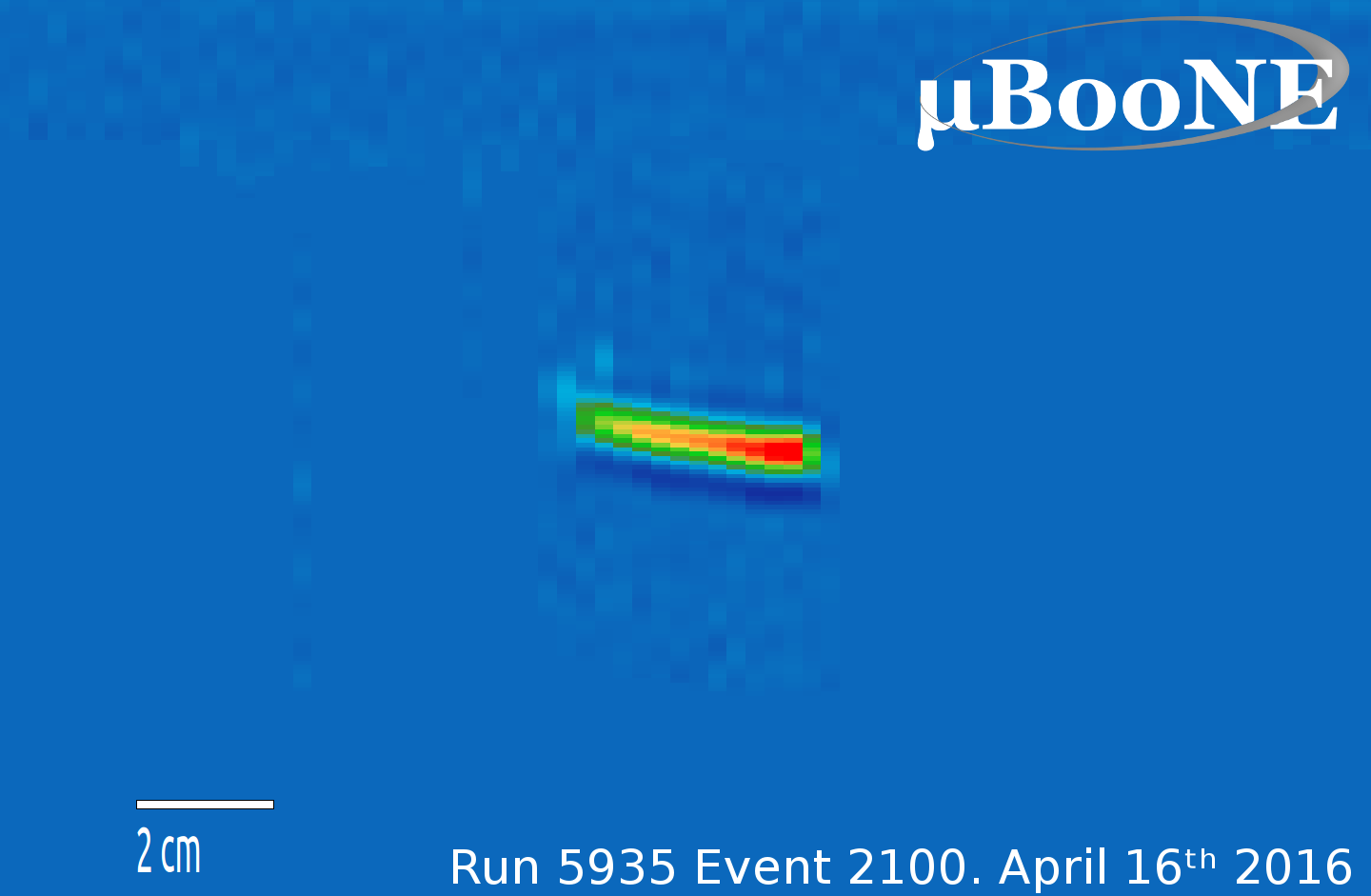}
  \caption{ Proton track candidate in MicroBooNE data. The track was selected
  by the decision tree classifier as being very likely a proton.
  \label{fig:evdnc}}
\end{figure}

\begin{figure}
  \includegraphics[scale=0.225]{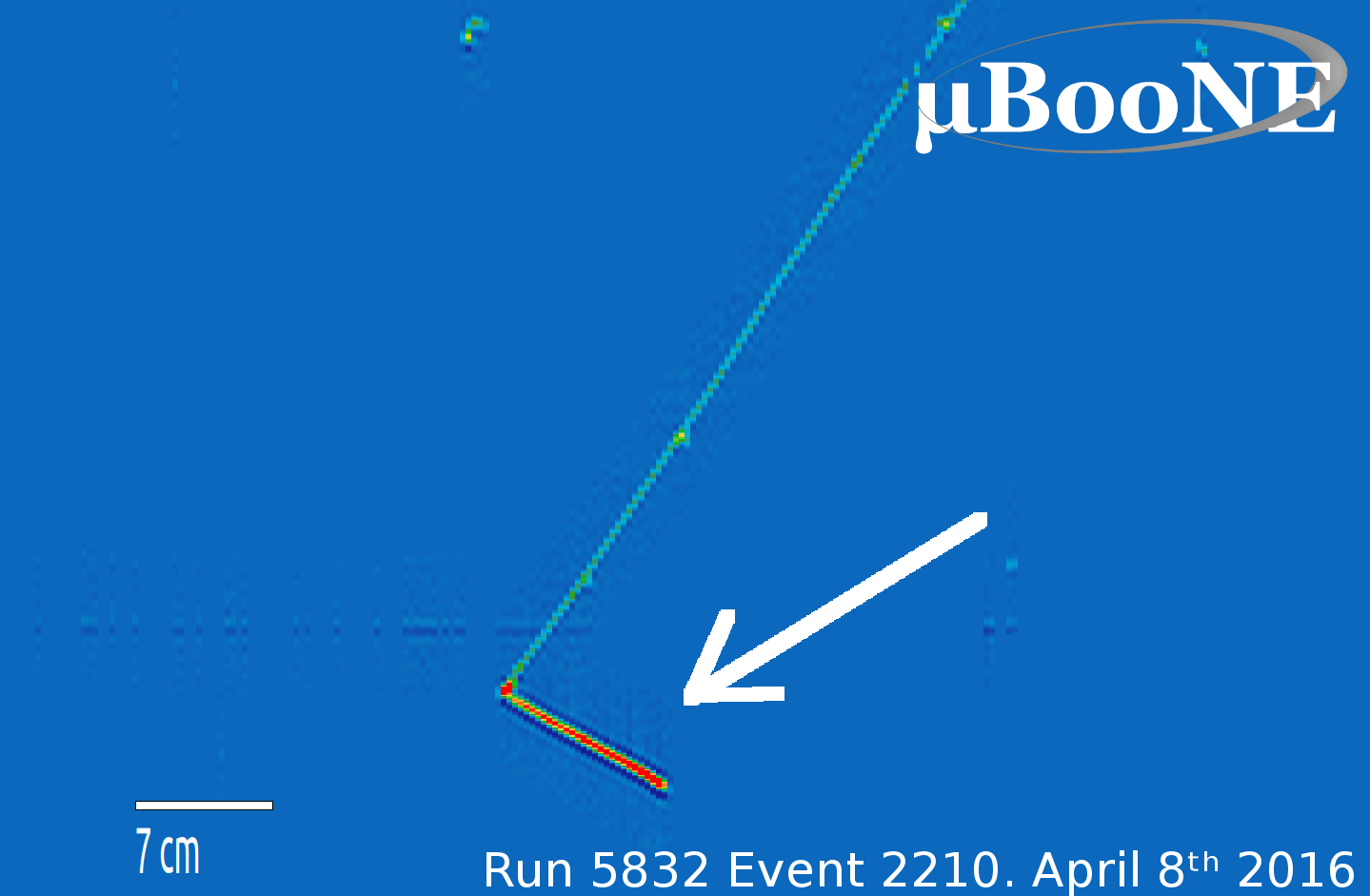}
  \caption{ Proton track candidate in MicroBooNE data. The white arrow points
  to a track that was selected by the decision tree classifier as being very
  likely a proton. \label{fig:evdcc}}
\end{figure}

The decision tree classifier was used on a small sample of MicroBooNE data as a
performance check. Figures~\ref{fig:evdcc}~and~\ref{fig:evdnc} show tracks from
the data sample that were selected by the classifier as being very likely
protons.

\subsection{NC elastic event selection}
So far, we have kept the proton selection general to all interaction types. For
NC elastic events, we would use the output of the decision trees along with
other event information such as the total number of reconstructed tracks to select
the events of interest. This can also be used to select charged-current elastic
events with a similar efficiency to use for normalization of the NC elastic
cross section. If we are only interested in one specific topology, and do not
wish to be general, it is trivial to re-train the classifier using protons from
NC elastic interactions as the only positive input and protons from other
interactions as a background input.

\section{Conclusions \label{conclusion}}
Whether the strange quarks in the nucleon sea contribute negatively or not at
all to the spin of the nucleon is an open question. Elastic neutrino-proton
scattering offers an unique way to determine $\Delta s$ that is independent of
the assumptions required by previous measurements. The MicroBooNE liquid argon
TPC can detect low-$Q^2$ NC elastic events and is currently taking neutrino
data at Fermilab. Automated event reconstruction and selection methods are
being developed to analyze the large amount of high-resolution neutrino events
in MicroBooNE.

\begin{acknowledgments}
This work was supported by the US Department of Energy, Office of Science,
Medium Energy Nuclear Physics Program.
\end{acknowledgments}

\bibliography{woodruff_spin2016_proceedings_arxiv.bib}

\providecommand{\noopsort}[1]{}\providecommand{\singleletter}[1]{#1}%
\begin{thebibliography}{21}%
\makeatletter
\providecommand \@ifxundefined [1]{%
 \@ifx{#1\undefined}
}%
\providecommand \@ifnum [1]{%
 \ifnum #1\expandafter \@firstoftwo
 \else \expandafter \@secondoftwo
 \fi
}%
\providecommand \@ifx [1]{%
 \ifx #1\expandafter \@firstoftwo
 \else \expandafter \@secondoftwo
 \fi
}%
\providecommand \natexlab [1]{#1}%
\providecommand \enquote  [1]{``#1''}%
\providecommand \bibnamefont  [1]{#1}%
\providecommand \bibfnamefont [1]{#1}%
\providecommand \citenamefont [1]{#1}%
\providecommand \href@noop [0]{\@secondoftwo}%
\providecommand \href [0]{\begingroup \@sanitize@url \@href}%
\providecommand \@href[1]{\@@startlink{#1}\@@href}%
\providecommand \@@href[1]{\endgroup#1\@@endlink}%
\providecommand \@sanitize@url [0]{\catcode `\\12\catcode `\$12\catcode
  `\&12\catcode `\#12\catcode `\^12\catcode `\_12\catcode `\%12\relax}%
\providecommand \@@startlink[1]{}%
\providecommand \@@endlink[0]{}%
\providecommand \url  [0]{\begingroup\@sanitize@url \@url }%
\providecommand \@url [1]{\endgroup\@href {#1}{\urlprefix }}%
\providecommand \urlprefix  [0]{URL }%
\providecommand \Eprint [0]{\href }%
\providecommand \doibase [0]{http://dx.doi.org/}%
\providecommand \selectlanguage [0]{\@gobble}%
\providecommand \bibinfo  [0]{\@secondoftwo}%
\providecommand \bibfield  [0]{\@secondoftwo}%
\providecommand \translation [1]{[#1]}%
\providecommand \BibitemOpen [0]{}%
\providecommand \bibitemStop [0]{}%
\providecommand \bibitemNoStop [0]{.\EOS\space}%
\providecommand \EOS [0]{\spacefactor3000\relax}%
\providecommand \BibitemShut  [1]{\csname bibitem#1\endcsname}%
\let\auto@bib@innerbib\@empty
\bibitem [{\citenamefont {Alberico}\ \emph {et~al.}(2002)\citenamefont
  {Alberico}, \citenamefont {Bilenky},\ and\ \citenamefont
  {Maieron}}]{Alberico01}%
  \BibitemOpen
  \bibfield  {author} {\bibinfo {author} {\bibfnamefont {W.~M.}\ \bibnamefont
  {Alberico}}, \bibinfo {author} {\bibfnamefont {S.~M.}\ \bibnamefont
  {Bilenky}}, \ and\ \bibinfo {author} {\bibfnamefont {C.}~\bibnamefont
  {Maieron}},\ }\href {\doibase 10.1016/S0370-1573(01)00058-8} {\bibfield
  {journal} {\bibinfo  {journal} {Phys. Rept.}\ }\textbf {\bibinfo {volume}
  {358}},\ \bibinfo {pages} {227} (\bibinfo {year} {2002})},\ \Eprint
  {http://arxiv.org/abs/hep-ph/0102269} {arXiv:hep-ph/0102269 [hep-ph]}
  \BibitemShut {NoStop}%
\bibitem [{\citenamefont {Ashman}\ \emph {et~al.}(1989)\citenamefont {Ashman}
  \emph {et~al.}}]{Ashman89}%
  \BibitemOpen
  \bibfield  {author} {\bibinfo {author} {\bibfnamefont {J.}~\bibnamefont
  {Ashman}} \emph {et~al.} (\bibinfo {collaboration} {European Muon}),\
  }\bibfield  {booktitle} {\emph {\bibinfo {booktitle} {{Internal spin
  structure of the nucleon. Proceedings, Symposium, SMC Meeting, New Haven,
  USA, January 5-6, 1994}}},\ }\href {\doibase 10.1016/0550-3213(89)90089-8}
  {\bibfield  {journal} {\bibinfo  {journal} {Nucl. Phys.}\ }\textbf {\bibinfo
  {volume} {B328}},\ \bibinfo {pages} {1} (\bibinfo {year} {1989})}\BibitemShut
  {NoStop}%
\bibitem [{\citenamefont {{Ellis, John and Jaffe, Robert}}(1974)}]{Ellis74}%
  \BibitemOpen
  \bibfield  {author} {\bibinfo {author} {\bibnamefont {{Ellis, John and Jaffe,
  Robert}}},\ }\href {\doibase 10.1103/PhysRevD.9.1444} {\bibfield  {journal}
  {\bibinfo  {journal} {Phys. Rev. D}\ }\textbf {\bibinfo {volume} {9}},\
  \bibinfo {pages} {1444} (\bibinfo {year} {1974})}\BibitemShut {NoStop}%
\bibitem [{\citenamefont {Aidala}\ \emph {et~al.}(2013)\citenamefont {Aidala},
  \citenamefont {Bass}, \citenamefont {Hasch},\ and\ \citenamefont
  {Mallot}}]{Aidala12}%
  \BibitemOpen
  \bibfield  {author} {\bibinfo {author} {\bibfnamefont {C.~A.}\ \bibnamefont
  {Aidala}}, \bibinfo {author} {\bibfnamefont {S.~D.}\ \bibnamefont {Bass}},
  \bibinfo {author} {\bibfnamefont {D.}~\bibnamefont {Hasch}}, \ and\ \bibinfo
  {author} {\bibfnamefont {G.~K.}\ \bibnamefont {Mallot}},\ }\href {\doibase
  10.1103/RevModPhys.85.655} {\bibfield  {journal} {\bibinfo  {journal} {Rev.
  Mod. Phys.}\ }\textbf {\bibinfo {volume} {85}},\ \bibinfo {pages} {655}
  (\bibinfo {year} {2013})},\ \Eprint {http://arxiv.org/abs/1209.2803}
  {arXiv:1209.2803 [hep-ph]} \BibitemShut {NoStop}%
\bibitem [{\citenamefont {Armstrong}\ and\ \citenamefont
  {McKeown}(2012)}]{Armstrong12}%
  \BibitemOpen
  \bibfield  {author} {\bibinfo {author} {\bibfnamefont {D.~S.}\ \bibnamefont
  {Armstrong}}\ and\ \bibinfo {author} {\bibfnamefont {R.~D.}\ \bibnamefont
  {McKeown}},\ }\href {\doibase 10.1146/annurev-nucl-102010-130419} {\bibfield
  {journal} {\bibinfo  {journal} {Ann. Rev. Nucl. Part. Sci.}\ }\textbf
  {\bibinfo {volume} {62}},\ \bibinfo {pages} {337} (\bibinfo {year} {2012})},\
  \Eprint {http://arxiv.org/abs/1207.5238} {arXiv:1207.5238 [nucl-ex]}
  \BibitemShut {NoStop}%
\bibitem [{\citenamefont {Cates}\ \emph {et~al.}(2011)\citenamefont {Cates},
  \citenamefont {de~Jager}, \citenamefont {Riordan},\ and\ \citenamefont
  {Wojtsekhowski}}]{Cates11}%
  \BibitemOpen
  \bibfield  {author} {\bibinfo {author} {\bibfnamefont {G.~D.}\ \bibnamefont
  {Cates}}, \bibinfo {author} {\bibfnamefont {C.~W.}\ \bibnamefont {de~Jager}},
  \bibinfo {author} {\bibfnamefont {S.}~\bibnamefont {Riordan}}, \ and\
  \bibinfo {author} {\bibfnamefont {B.}~\bibnamefont {Wojtsekhowski}},\ }\href
  {\doibase 10.1103/PhysRevLett.106.252003} {\bibfield  {journal} {\bibinfo
  {journal} {Phys. Rev. Lett.}\ }\textbf {\bibinfo {volume} {106}},\ \bibinfo
  {pages} {252003} (\bibinfo {year} {2011})},\ \Eprint
  {http://arxiv.org/abs/1103.1808} {arXiv:1103.1808 [nucl-ex]} \BibitemShut
  {NoStop}%
\bibitem [{\citenamefont {Patrignani}\ \emph {et~al.}(2016)\citenamefont
  {Patrignani} \emph {et~al.}}]{Olive16}%
  \BibitemOpen
  \bibfield  {author} {\bibinfo {author} {\bibfnamefont {C.}~\bibnamefont
  {Patrignani}} \emph {et~al.} (\bibinfo {collaboration} {Particle Data
  Group}),\ }\href {\doibase 10.1088/1674-1137/40/10/100001} {\bibfield
  {journal} {\bibinfo  {journal} {Chin. Phys.}\ }\textbf {\bibinfo {volume}
  {C40}},\ \bibinfo {pages} {100001} (\bibinfo {year} {2016})}\BibitemShut
  {NoStop}%
\bibitem [{\citenamefont {Ahrens}\ \emph {et~al.}(1987)\citenamefont {Ahrens}
  \emph {et~al.}}]{Ahrens86}%
  \BibitemOpen
  \bibfield  {author} {\bibinfo {author} {\bibfnamefont {L.~A.}\ \bibnamefont
  {Ahrens}} \emph {et~al.},\ }\href {\doibase 10.1103/PhysRevD.35.785}
  {\bibfield  {journal} {\bibinfo  {journal} {Phys. Rev.}\ }\textbf {\bibinfo
  {volume} {D35}},\ \bibinfo {pages} {785} (\bibinfo {year}
  {1987})}\BibitemShut {NoStop}%
\bibitem [{\citenamefont {{Garvey, G. T. and Louis, W. C. and White, D.
  H.}}(1993)}]{Garvey93}%
  \BibitemOpen
  \bibfield  {author} {\bibinfo {author} {\bibnamefont {{Garvey, G. T. and
  Louis, W. C. and White, D. H.}}},\ }\href {\doibase 10.1103/PhysRevC.48.761}
  {\bibfield  {journal} {\bibinfo  {journal} {Phys. Rev. C}\ }\textbf {\bibinfo
  {volume} {48}},\ \bibinfo {pages} {761} (\bibinfo {year} {1993})}\BibitemShut
  {NoStop}%
\bibitem [{\citenamefont {Aguilar-Arevalo}\ \emph {et~al.}(2010)\citenamefont
  {Aguilar-Arevalo} \emph {et~al.}}]{Aguilar10}%
  \BibitemOpen
  \bibfield  {author} {\bibinfo {author} {\bibfnamefont {A.~A.}\ \bibnamefont
  {Aguilar-Arevalo}} \emph {et~al.} (\bibinfo {collaboration} {MiniBooNE}),\
  }\href {\doibase 10.1103/PhysRevD.82.092005} {\bibfield  {journal} {\bibinfo
  {journal} {Phys. Rev.}\ }\textbf {\bibinfo {volume} {D82}},\ \bibinfo {pages}
  {092005} (\bibinfo {year} {2010})},\ \Eprint {http://arxiv.org/abs/1007.4730}
  {arXiv:1007.4730 [hep-ex]} \BibitemShut {NoStop}%
\bibitem [{\citenamefont {Acciarri}\ \emph {et~al.}(2016)\citenamefont
  {Acciarri} \emph {et~al.}}]{UBdetector}%
  \BibitemOpen
  \bibfield  {author} {\bibinfo {author} {\bibfnamefont {R.}~\bibnamefont
  {Acciarri}} \emph {et~al.} (\bibinfo {collaboration} {MicroBooNE}),\
  }\href@noop {} {\bibfield  {journal} {\bibinfo  {journal} {Submitted to:
  JINST}\ } (\bibinfo {year} {2016})},\ \Eprint
  {http://arxiv.org/abs/1612.05824} {arXiv:1612.05824 [physics.ins-det]}
  \BibitemShut {NoStop}%
\bibitem [{\citenamefont {Aguilar-Arevalo}\ \emph {et~al.}(2009)\citenamefont
  {Aguilar-Arevalo} \emph {et~al.}}]{Aguilar09}%
  \BibitemOpen
  \bibfield  {author} {\bibinfo {author} {\bibfnamefont {A.~A.}\ \bibnamefont
  {Aguilar-Arevalo}} \emph {et~al.} (\bibinfo {collaboration} {MiniBooNE}),\
  }\href {\doibase 10.1103/PhysRevD.79.072002} {\bibfield  {journal} {\bibinfo
  {journal} {Phys. Rev.}\ }\textbf {\bibinfo {volume} {D79}},\ \bibinfo {pages}
  {072002} (\bibinfo {year} {2009})},\ \Eprint {http://arxiv.org/abs/0806.1449}
  {arXiv:0806.1449 [hep-ex]} \BibitemShut {NoStop}%
\bibitem [{\citenamefont {Pate}\ and\ \citenamefont {Trujillo}(2014)}]{Pate13}%
  \BibitemOpen
  \bibfield  {author} {\bibinfo {author} {\bibfnamefont {S.}~\bibnamefont
  {Pate}}\ and\ \bibinfo {author} {\bibfnamefont {D.}~\bibnamefont
  {Trujillo}},\ }\bibfield  {booktitle} {\emph {\bibinfo {booktitle}
  {{Proceedings, 25th International Nuclear Physics Conference (INPC 2013):
  Florence, Italy, June 2-7, 2013}}},\ }\href {\doibase
  10.1051/epjconf/20146606018} {\bibfield  {journal} {\bibinfo  {journal} {EPJ
  Web Conf.}\ }\textbf {\bibinfo {volume} {66}},\ \bibinfo {pages} {06018}
  (\bibinfo {year} {2014})},\ \Eprint {http://arxiv.org/abs/1308.5694}
  {arXiv:1308.5694 [hep-ph]} \BibitemShut {NoStop}%
\bibitem [{\citenamefont {{The MicroBooNE
  Collaboration}}(2016{\natexlab{a}})}]{UBcosmic}%
  \BibitemOpen
  \bibfield  {author} {\bibinfo {author} {\bibnamefont {{The MicroBooNE
  Collaboration}}},\ }\href@noop {} {\bibfield  {journal} {\bibinfo  {journal}
  {MicroBooNE Public Note}\ } (\bibinfo {year}
  {2016}{\natexlab{a}})}\BibitemShut {NoStop}%
{ MICROBOONE-NOTE-1002-PUB}
\bibitem [{\citenamefont {{The MicroBooNE
  Collaboration}}(2016{\natexlab{b}})}]{UBpandora}%
  \BibitemOpen
  \bibfield  {author} {\bibinfo {author} {\bibnamefont {{The MicroBooNE
  Collaboration}}},\ }\href@noop {} {\bibfield  {journal} {\bibinfo  {journal}
  {MicroBooNE Public Note}\ } (\bibinfo {year}
  {2016}{\natexlab{b}})}\BibitemShut {NoStop}%
{ MICROBOONE-NOTE-1015-PUB}
\bibitem [{\citenamefont {Church}(2013)}]{Church13}%
  \BibitemOpen
  \bibfield  {author} {\bibinfo {author} {\bibfnamefont {E.~D.}\ \bibnamefont
  {Church}},\ }\href@noop {} {\  (\bibinfo {year} {2013})},\ \Eprint
  {http://arxiv.org/abs/1311.6774} {arXiv:1311.6774 [physics.ins-det]}
  \BibitemShut {NoStop}%
\bibitem [{\citenamefont {{The MicroBooNE
  Collaboration}}(2016{\natexlab{c}})}]{UBnoise}%
  \BibitemOpen
  \bibfield  {author} {\bibinfo {author} {\bibnamefont {{The MicroBooNE
  Collaboration}}},\ }\href@noop {} {\bibfield  {journal} {\bibinfo  {journal}
  {MicroBooNE Public Note}\ } (\bibinfo {year}
  {2016}{\natexlab{c}})}\BibitemShut {NoStop}%
{ MICROBOONE-NOTE-1009-PUB}
\bibitem [{\citenamefont {{Tianqi Chen and Carlos Guestrin}}(2016)}]{Chen16}%
  \BibitemOpen
  \bibfield  {author} {\bibinfo {author} {\bibnamefont {{Tianqi Chen and Carlos
  Guestrin}}},\ }\href {http://arxiv.org/abs/1603.02754} {\bibfield  {journal}
  {\bibinfo  {journal} {CoRR}\ } (\bibinfo {year} {2016})},\ \Eprint
  {http://arxiv.org/abs/1603.02754} {arXiv:1603.02754} \BibitemShut {NoStop}%
\bibitem [{\citenamefont {Andreopoulos}\ \emph {et~al.}(2010)\citenamefont
  {Andreopoulos} \emph {et~al.}}]{Andreopoulos09}%
  \BibitemOpen
  \bibfield  {author} {\bibinfo {author} {\bibfnamefont {C.}~\bibnamefont
  {Andreopoulos}} \emph {et~al.},\ }\href {\doibase 10.1016/j.nima.2009.12.009}
  {\bibfield  {journal} {\bibinfo  {journal} {Nucl. Instrum. Meth.}\ }\textbf
  {\bibinfo {volume} {A614}},\ \bibinfo {pages} {87} (\bibinfo {year}
  {2010})},\ \Eprint {http://arxiv.org/abs/0905.2517} {arXiv:0905.2517
  [hep-ph]} \BibitemShut {NoStop}%
\bibitem [{\citenamefont {Heck}\ \emph {et~al.}(1998)\citenamefont {Heck},
  \citenamefont {Schatz}, \citenamefont {Thouw}, \citenamefont {Knapp},\ and\
  \citenamefont {Capdevielle}}]{Heck98}%
  \BibitemOpen
  \bibfield  {author} {\bibinfo {author} {\bibfnamefont {D.}~\bibnamefont
  {Heck}}, \bibinfo {author} {\bibfnamefont {G.}~\bibnamefont {Schatz}},
  \bibinfo {author} {\bibfnamefont {T.}~\bibnamefont {Thouw}}, \bibinfo
  {author} {\bibfnamefont {J.}~\bibnamefont {Knapp}}, \ and\ \bibinfo {author}
  {\bibfnamefont {J.~N.}\ \bibnamefont {Capdevielle}},\ }\href@noop {} {\
  (\bibinfo {year} {1998})}\BibitemShut {NoStop}%
\bibitem [{\citenamefont {Agostinelli}\ \emph {et~al.}(2003)\citenamefont
  {Agostinelli} \emph {et~al.}}]{Agostinelli02}%
  \BibitemOpen
  \bibfield  {author} {\bibinfo {author} {\bibfnamefont {S.}~\bibnamefont
  {Agostinelli}} \emph {et~al.} (\bibinfo {collaboration} {GEANT4}),\ }\href
  {\doibase 10.1016/S0168-9002(03)01368-8} {\bibfield  {journal} {\bibinfo
  {journal} {Nucl. Instrum. Meth.}\ }\textbf {\bibinfo {volume} {A506}},\
  \bibinfo {pages} {250} (\bibinfo {year} {2003})}\BibitemShut {NoStop}%
\end{thebibliography}%

\end{document}